\documentclass[%
 reprint,
 amsmath,amssymb,
aps,prb,
]{revtex4-1}

\usepackage{graphicx}
\usepackage{dcolumn}
\usepackage{bm}
\usepackage{textcomp, gensymb}
\usepackage{lmodern}
\newcolumntype{M}[1]{>{\centering\arraybackslash}m{#1}}

\begin{document}

\preprint{APS/123-QED}

\title{Coexistence of bulk and surface states probed by Shubnikov--de Haas oscillations in Bi$_2$Se$_3$ with high charge-carrier density}

\author{E. K. de Vries$^1$}
\email{eric.de.vries@rug.nl}	
\author{S. Pezzini$^2$}
\author{M. J. Meijer$^2$}
\author{N. Koirala$^3$}
\author{M. Salehi$^4$}
\author{J. Moon$^3$}
\author{S. Oh$^3$}
\author{S. Wiedmann$^2$}
\author{T. Banerjee$^1$}
	\email{t.banerjee@rug.nl}
\affiliation{$^1$University of Groningen, Zernike Institute for Advanced Materials, 9747 AG Groningen, The Netherlands}
\affiliation{$^2$High Field Magnet Laboratory (HFML-EMFL) and Institute for Molecules and Materials, Radboud University, 6525 ED Nijmegen, The Netherlands}
\affiliation{$^3$Department of Physics $\&$ Astronomy, Rutgers, The State University of New Jersey, Piscataway, New Jersey 08854, USA}
\affiliation{$^4$Department of Materials Science $\&$ Engineering, Rutgers University, Piscataway, New Jersey 08854, USA}

\begin{abstract}
Topological insulators are ideally represented as having an insulating bulk with topologically protected, spin-textured surface states. However, it is increasingly becoming clear that these surface transport channels can be accompanied by a finite conducting bulk, as well as additional topologically trivial surface states. To investigate these parallel conduction transport channels, we studied Shubnikov--de Haas oscillations in Bi$_2$Se$_3$ thin films, in high magnetic fields up to 30 T so as to access channels with a lower mobility. We identify a clear Zeeman-split bulk contribution to the oscillations from a comparison between the charge-carrier densities extracted from the magnetoresistance and the oscillations. Furthermore, our analyses indicate the presence of a two-dimensional state and signatures of additional states the origin of which cannot be conclusively determined. Our findings underpin the necessity of theoretical studies on the origin of and the interplay between these parallel conduction channels for a careful analysis of the material's performance.

\end{abstract}

\maketitle

Topological insulators (TIs), hosting spin-momentum locked surface states, received considerable interest in the past decade potentially serving as a platform for exploring many interesting concepts in physics\cite{hasan_colloquium:_2010,ando_topological_2013,moore_birth_2010}. These surface states have been well investigated by surface sensitive techniques like (spin-)angular resolved photo-emission spectroscopy\cite{xia_observation_2009,hsieh_tunable_2009} and scanning tunneling microscopy and spectroscopy\cite{alpichshev_stm_2010,roushan_topological_2009,hanaguri_momentum-resolved_2010}. Such techniques adequately describe the electronic properties of the (non)trivial surface states, but cannot account for additional transport features as observed in (magneto)transport experiments. To employ topological insulators in solid-state devices, direct access and understanding of these additional surface states in transport experiments are needed. 
  
Studying Shubnikov--de Haas (SdH) oscillations can reveal the existence of such surface states where parameters like mobility, charge-carrier density, the dimensionality and the Berry phase of the states can be determined\cite{ando_topological_2013}. Earlier studies on various Bi-based topological insulators report on single or double frequency SdH oscillations\cite{kohler_conduction_1973,kulbachinskii_conduction-band_1999,analytis_bulk_2010,butch_strong_2010,ren_large_2010,petrushevsky_probing_2012,lawson_quantum_2012,bansal_thickness-independent_2012,fang_catalyst-free_2012,qu_coexistence_2013,shrestha_shubnikovchar21haas_2014,devidas_role_2014,zhang_observations_2014,piot_hole_2016}, where it is often claimed that these oscillations originate from the top and bottom topological surface state (TSS) with the expected Berry phase and angular dependence. The magnetic field strength used in these studies is usually up to 15 T, which can only probe transport channels with a relatively high mobility, whereas nonlinear Hall measurements indicate additional channels with a lower mobility to be present. Besides a finite conducting bulk, these additional (topologically trivial) channels can originate from variations in the electrostatic potential near the surfaces and can also be spin textured \cite{bianchi_coexistence_2010,king_large_2011,bahramy_emergent_2012}, which we will refer to as two-dimensional electron gas (2DEG). From earlier transport measurements, the mobilities of the different channels are found to be on the order of 50--500 and $\raise.17ex\hbox{$\scriptstyle\sim$}$3000 cm$^2$(V s)$^{-1}$ where the low mobility channel has a higher charge-carrier density\cite{bansal_thickness-independent_2012,park_terahertz_2015}. Notably, from these numbers one can find that in terms of conductivity these channels can contribute equally to the electrical transport.  

Motivated by these works, we performed magnetotransport experiments up to 30 T and studied SdH oscillations to explore the most prominent conduction channels and additional channels with mobilities below 1000 cm$^2$(V s)$^{-1}$ ($\textit{\micro}B\gg1$, where $\textit{\micro}$ is the charge mobility and $B$ the applied magnetic field strength). The magnetotransport is studied in thin films of Bi$_2$Se$_3$ so as to minimize bulk effects and amplify the topologically trivial and nontrivial surface states. In contrast to earlier works, we will show that the bulk channel with a high mobility is present along with a prominent two-dimensional (2D) channel which can be linked to the topological surface states. Our findings indicate the presence of additional channels with a lower mobility that cannot be precisely resolved from the oscillations. Similar to Ref.~\cite{syers_ambipolar_2017}, we compare charge-carrier densities from the SdH oscillations as well as from the magnetoresistance and study the dimensionality of the various channels in order to unravel the origin of these states.

\begin{figure}[!h]
	\centering
		\includegraphics[width=\columnwidth]{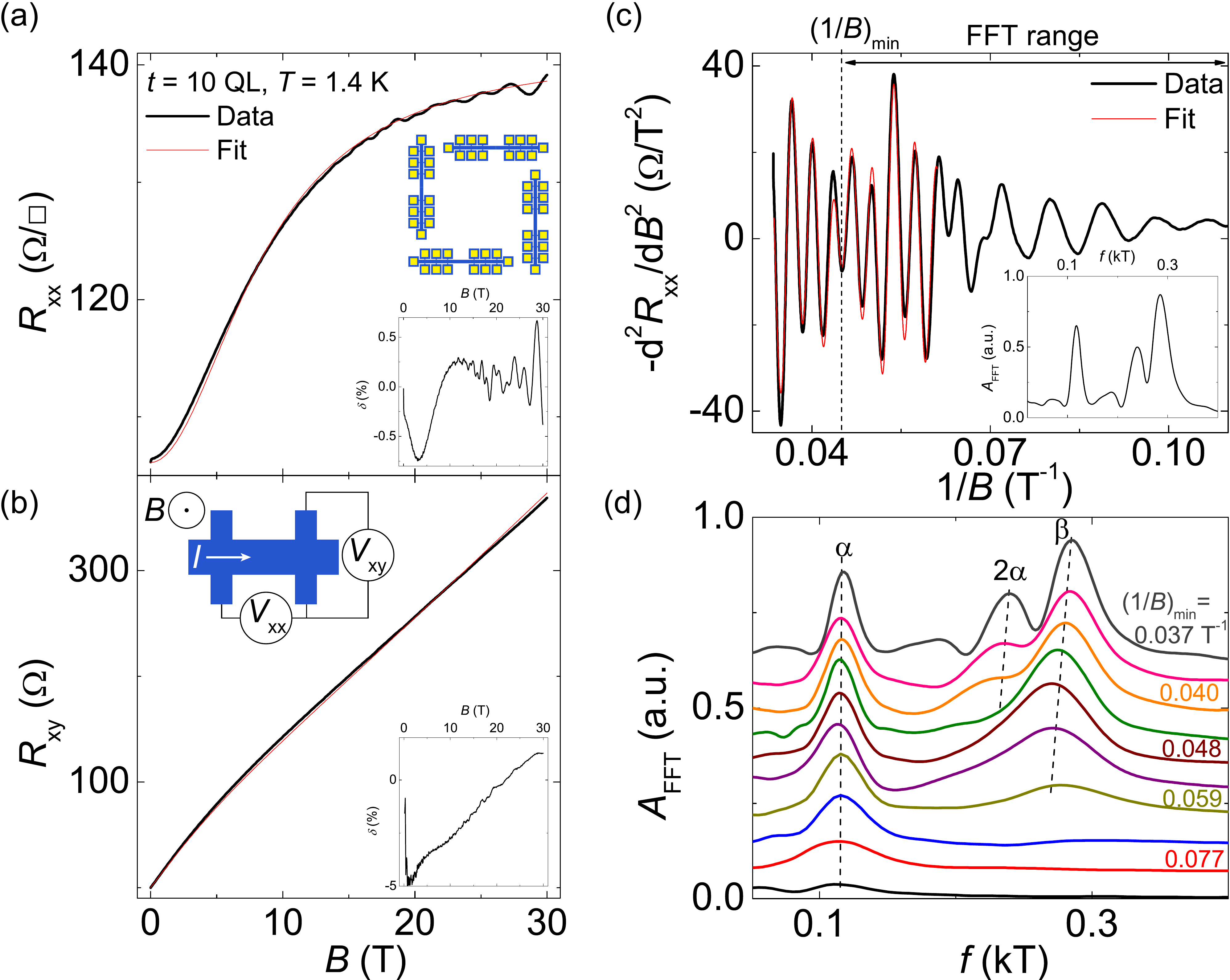}
		\caption{(a) Out-of-plane magnetic field dependence of the longitudinal sheet resistance $R_{\textup{xx}}$ and (b) the Hall resistance $R_{\textup{xy}}$ for $t$ = 10 QL at $T$ = 1.4 K. The data (black) can be fitted with the two channel model (red) in good agreement. Oscillations in $R_{\textup{xx}}$ are clearly visible beyond 15 T. Insets (a): Residual $\delta$ vs magnetic field and Hall bar geometry with TI channel in blue and contact pads in yellow. Inset (b): Residual $\delta$ vs magnetic field. (c) The second derivative of the resistance with respect to the magnetic field --$d^2R_{\textup{xx}}$/$dB^2$ plotted vs 1/$B$. A clear oscillatory pattern is present with multiple oscillations. Beyond 15.5 T (0.065 T$^{-1}$) the oscillatory pattern can be reconstructed from oscillations with $f_{\alpha}$ = 0.122$\pm$0.003 kT, $f_{\textup{2}\alpha} = $0.236$\pm$0.003 kT, and $f_{\beta}$ = 0.291$\pm$0.001 kT. Inset: Resulting FFT spectrum of the full oscillatory pattern. (d) Magnetic field evolution of the (smoothed) FFT spectrum analyzed for different FFT ranges starting from 11 T towards higher fields with steps of 2 T, as schematically depicted in (c). The FFT amplitude $A_{\textup{FFT}}$ is plotted vs frequency $f$ where the curves are offsetted by 0.07 for clarity.}
	\label{fig:MRandOscillations}
\end{figure}

In this paper, we used thin films of n-type Bi$_2$Se$_3$ with thickness $t$ = 10, 20, 30, and 100 quintuple layers (QL) grown by molecular-beam epitaxy (MBE) on Al$_2$O$_3$(0001) substrates in a custom-designed SVTA MOS-V-2 MBE system at a base pressure lower than 5$\times$10$^{-10}$ Torr following the methods as described in previous work\cite{bansal_epitaxial_2011}. The quality of the obtained films was characterized through various techniques\cite{bansal_thickness-independent_2012,valdes_aguilar_terahertz_2012,de_vries_towards_2015,dai_restoring_2015,brahlek_disorder-driven_2016}. The films were patterned into Hall bars by using a combination of photolithography and Ar plasma dry etching. Contact pads consisting of Ti(5)/Au(70 nm) were made by combining photolithography with electron-beam evaporation. The resulting Hall bars (inset Fig.~\ref{fig:MRandOscillations}a) have dimensions of 2400$\times$100 $\micro$m$^2$ where the resistance is measured over a probing length between 1400 and 2000 $\micro$m. The magnetotransport measurements have been performed in a cryostat with an out-of-plane rotation stage placed in a 30 T Bitter-type magnet in a four-probe geometry using the ac modulation technique at an ac current bias of 1 $\micro$A. In this paper, mainly the results on the sample with $t$ = 10 QL will be discussed and comparisons will be made to the samples with larger thickness. Most of the results of the thicker samples can be found in the Supplemental Material below.

The typical out-of-plane magnetic field dependence of the longitudinal sheet resistance $R_{\textup{xx}}$ measured for the sample with $t$ = 10 QL is shown in Fig.~\ref{fig:MRandOscillations}a where $R_{\textup{xx}}$ tends to saturate at high magnetic fields. From this data and those for larger thickness as well as from the fitting (see below), we observe that the order of saturation is determined by the low mobility channel and the parabolic response at low fields is governed by the high mobility channel. Furthermore, the presence of at least two channels is clear from the nonlinear Hall resistance $R_{\textup{xy}}$ [Fig.~\ref{fig:MRandOscillations}b]. As also shown earlier\cite{de_vries_towards_2015}, we observe a slight upturn with a change in $R_{\textup{xx}}$ $\raise.17ex\hbox{$\scriptstyle\sim$}$0.2$\%$ for samples with $t$ = 10--30 QL below 10 K, indicative of the presence of defect states\cite{butch_strong_2010,kulbachinskii_conduction-band_1999,kohler_galvanomagnetic_1975}. From the out-of-plane field dependence of the longitudinal and transverse resistance $R_{\textup{xx}}(B)$ and $R_{\textup{xy}}(B)$, we can extract the sheet carrier density $n_i$ and mobility $\mu_i$ for only two channels, which we expect to be due to the bulk and surface state(s). For that, we use a semi-classical Drude model where contributions from two parallel channels are summed in the conductivity tensor $\hat{\sigma}$, which relates to the resistivity as $\hat{\rho}$ = $\hat{\sigma}^{-1}$ (more details on the analysis can be found in the Supplemental Material below):

\begin{equation}
\label{eq:Drude}
{\sigma}_{\textup{xx}}=\frac{n_{1}e\mu_{1}}{1+\mu_1^{2}B^{2}}+\frac{n_{2}e\mu_{2}}{1+\mu_2^{2}B^{2}}, 
{\sigma}_{\textup{xy}}=\frac{n_{1}e\mu_{1}^{2}B}{1+\mu_1^{2}B^{2}}+\frac{n_{2}e\mu_{2}^{2}B}{1+\mu_2^{2}B^{2}}
\end{equation}

\noindent
As found from our analysis, simultaneous fitting of the $R_{\textup{xx}}$ and $R_{\textup{xy}}$ is required since $R_{\textup{xx}}$ has a strong effect on the mobility and therefore will change the values found for $n_i$ from $R_{\textup{xy}}$. An example of the simultaneous fit to the magnetoresistance curves $R_{\textup{xx}}$ and $R_{\textup{xy}}$ for the sample with $t$ = 10 QL at 1.4 K is displayed in Fig.~\ref{fig:MRandOscillations}a. A good agreement with the two-carrier model is obtained with a residual $\delta$ = ($R_{\textup{data}}$-$R_{\textup{fit}}$)/$R_{\textup{data}}$ between 1 and 5 $\%$ for both $R_{\textup{xx}}$ and $R_{\textup{xy}}$, but it is important to note that this analysis is limited to two channels and does not rule out the presence of more channels. Nevertheless, the good agreement between data and fit suggests that any additional state would have a similar mobility, which would add to an effective charge-carrier density in Eq.~(\ref{eq:Drude}).

An overview of the extracted charge-carrier properties for all film thicknesses can be found in Table~\ref{tab:DensityMobility}; the data and fits to the magnetoresistance for the samples with larger thickness can be found in the Supplemental Material below. We listed $n_{\textup{1}}/t$ because of its correspondence to the [three-dimensional (3D)] bulk channel (see discussion below), whereas $n_{\textup{2}}$ is most likely linked to a 2D channel. The model describes the magnetoresistance behavior for $t$ up to 30 QL very well, but deviations from the model are observed for $t$ = 100 QL, which will be discussed below. The correspondence between these extracted parameters and the information extracted from the SdH oscillations will be discussed in the remainder of this paper.

\begin{table}[b]
	\centering
	\resizebox{\columnwidth}{!}{
		\begin{tabular}{|M{1.3cm}|M{2.4cm}|M{2.1cm}|M{2cm}|M{2cm}|}
		\hline
		$t$(QL)&$n_{1}/t$($\times$10$^{19}$cm$^{-3}$)\rule{0pt}{2.6ex}&$n_{\textup{2}}$($\times$10$^{13}$cm$^{-2}$)\rule{0pt}{2.6ex}&$\mu_{1}$(cm$^{2}$(V s)$^{-1}$)\rule{0pt}{2.6ex}&$\mu_{2}$(cm$^{2}$(V s)$^{-1}$)\rule{0pt}{2.6ex}\\\hline
		10$\pm$1&1.8$\pm$0.1&3.4$\pm$0.2&2060$\pm$50&660$\pm$50\\\hline
		20$\pm$1&0.67$\pm$0.03&2.4$\pm$0.2&1190$\pm$50&290$\pm$50\\\hline
		30$\pm$1&0.49$\pm$0.02&2.7$\pm$0.1&1250$\pm$50&220$\pm$50\\\hline
		100$\pm$5&0.33$\pm$0.02&4.2$\pm$0.2&3800$\pm$100&500$\pm$100\\\hline			
		\end{tabular}}
	\caption{Overview of the extracted charge-carrier densities $n_{1}/t$, $n_2$ and mobilities $\textit{\micro}_1$, $\textit{\micro}_2$ from the magnetoresistance measurements using the two-carrier Drude model.}
	\label{tab:DensityMobility}
\end{table}

The possible presence of additional states can be analyzed by studying the SdH oscillations in $R_{\textup{xx}}$, provided that the mobility of the channels is high enough\cite{syers_ambipolar_2017}. For the sample with $t$ = 10 QL, these oscillations can be observed from $\raise.17ex\hbox{$\scriptstyle\sim$}$10 T onwards, which indicates that transport channels are present with a mobility on the order of 1000 cm$^2$(V s)$^{-1}$. This is in agreement with estimates for $\textit{\micro}_1$ as extracted from the magnetoresistance measurements (see Table~\ref{tab:DensityMobility}). To analyze the oscillations without the magnetoresistance background, the second derivative --$d^{2}R_{\textup{xx}}/$$dB^2$ is taken after interpolation and adjacent averaging of the data (see Supplemental Material below for more details). By plotting --$d^{2}R_{\textup{xx}}/$$dB^2$ versus 1/$B$, we find an oscillatory pattern that shows additional oscillations from 15 T ($\raise.17ex\hbox{$\scriptstyle\sim$}$0.067 T$^{-1}$, Fig.~\ref{fig:MRandOscillations}c). We can follow the development of the oscillations by looking at the evolution of the fast fourier transform (FFT) spectrum when taking different ranges starting from 9 T (\raise.17ex\hbox{$\scriptstyle\sim$}0.11 T$^{-1}$) towards higher fields, where lower mobility channels start to contribute [Fig.~\ref{fig:MRandOscillations}d]. Below 15 T, as depicted by the black, red, and blue line in Fig.~\ref{fig:MRandOscillations}d, one main frequency is observed indicated by $\alpha$ in the FFT spectrum as has been commonly reported in other works\cite{analytis_bulk_2010,butch_strong_2010,ren_large_2010,bansal_thickness-independent_2012,shrestha_shubnikovchar21haas_2014,devidas_role_2014,fang_catalyst-free_2012,lawson_quantum_2012}.

\begin{figure}[h]
	\centering
	\includegraphics[width=\columnwidth]{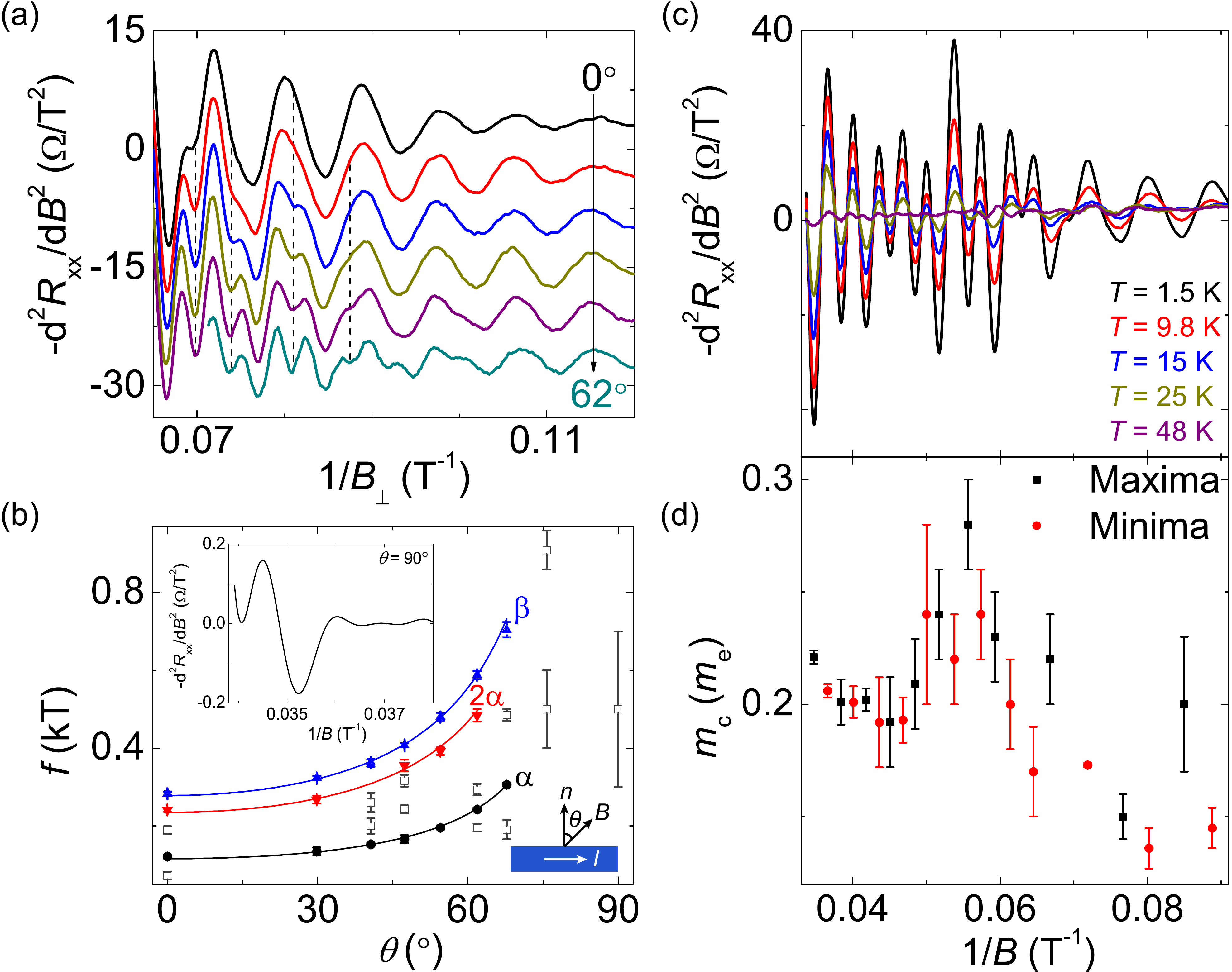}
	\caption{(a) $B_{\perp}$ dependence on the oscillations above 0.06 T$^{-1}$ for increasing angle $\theta$ as defined in the inset in (b). An enhancement of the oscillation amplitude is observed at locations indicated by the dashed lines. (b) Angular dependence of the frequencies extracted from the obtained FFT spectra. The dark gray, open square symbols designate additional peaks observed in the FFT spectra but which do not follow a clear angular dependent trend. The large error bars above 70$^{\circ}$ display the range of the peak position which cannot be determined accurately from the FFT spectrum. Insets: Remaining oscillation at $\theta$ = 90$^{\circ}$ and schematics of the relative directions of current $I$ and magnetic field $B$. (c) Temperature dependence of the oscillations measured from 1.5 to 48 K. (d) Extracted cyclotron masses per peak position following from the data in (c).}
	\label{fig:AngularAndTemperature}
\end{figure}

Beyond 15 T, we find the clear presence of the harmonic $2\alpha$ in the FFT spectrum which is due to the strong Zeeman splitting because of the large $g$ factor in this material \cite{kohler_g-factor_1975,wolos_$g$_2016,orlita_magneto-optics_2015,fauque_magnetothermoelectric_2013}. As shown in Fig.~\ref{fig:AngularAndTemperature}a, the occurrence of Zeeman splitting is justified by an enhancement in oscillation amplitude when studying its dependence on the perpendicular to the in-plane component of the magnetic field, $B_{\perp}$. Only the cyclotron energy is sensitive to this component, whereas the competing Zeeman energy is related to the total applied magnetic field. In addition to the high mobility channel linked to $\alpha$, we find a lower mobility channel denoted by $\beta$. The appearance of the oscillation linked to $\beta$ at higher magnetic field indicates that this channel has a mobility on the order of several times 100 cm$^2$(V s)$^{-1}$ which is in agreement with the lower value $\mu_{\textup{2}}$ found from the earlier analysis of the magnetoresistance (Table~\ref{tab:DensityMobility}). Using the extracted three frequencies, we can reconstruct the oscillatory pattern at high fields as shown in Fig.~\ref{fig:MRandOscillations}c with deviations in the peak amplitudes of the pattern. Due to the good agreement between data and the reconstructed oscillatory pattern, we can conclude that in the used magnetic field range the magnetotransport is dominated by these three frequencies. Nevertheless, additional channels with a lower mobility might be present but are beyond the resolution of our measurements. 

To explore the dimensionality of the observed conduction channels, we can look at the angular dependence of the magnetic field orientation on the position of the frequency peaks. For 2D states we expect that their frequencies $f$ scale with $f\propto$ 1/$\cos\theta$ where $\theta$ is the angle between the surface normal and the direction of the applied magnetic field (inset Fig.~\ref{fig:AngularAndTemperature}b). For bulk states it is commonly observed that $f(\theta)$ initially follows the similar behavior but saturates between 30 and 60$^{\circ}$, depending on the dimensions of the ellipsoidal pocket of these states\cite{kulbachinskii_conduction-band_1999,martin_bulk_2013, wiedmann_anisotropic_2016,piot_hole_2016,qu_coexistence_2013}. However, few earlier works\cite{analytis_bulk_2010, eto_angular-dependent_2010, lawson_quantum_2012} report on a similar 1/$\cos\theta$ dependence for the bulk states as well. Importantly, although we use thin films, the bulk will not show 2D behavior due to finite film thickness since the magnetic length $l_{\textup{B}}= \sqrt{\hbar/{eB}}\leq$ 8 nm at fields of 10 T from which we start observing the oscillations. From these considerations for $f(\theta)$ we can map out all observed peaks in the spectra at every angle $\theta$ and check whether they fit into a 2D or 3D picture. In Fig.~\ref{fig:AngularAndTemperature}b, the angular dependence of the observed peaks for $t$ = 10 QL is plotted from where we can trace the different channels $\alpha$, $2\alpha$, and $\beta$ up to an angle of 68$^{\circ}$. Beyond this angle, the resolution of separate spectral peaks is limited which is most probably linked to the low mobility of the channels and is manifested as a strong weakening of the oscillations at higher angles. Nevertheless, partially due to a higher mobility of the channel, we find a minor oscillation with $f$ = (0.5$\pm$0.2) kT at $\theta$ = 90$^{\circ}$ indicating that $f_{\alpha}$ (and its harmonic) saturates and is due to the bulk channel with an elongated Fermi pocket. For the $\beta$ peak, we find a 1/$\cos\theta$ behavior which can be linked to the appearance of a 2D state. 

Another way of clarifying the origin of the states is to extract the cyclotron mass from the temperature dependence of the oscillations which are observable up to $\raise.17ex\hbox{$\scriptstyle\sim$}$50 K as shown in Fig.~\ref{fig:AngularAndTemperature}c. Because of the presence of multiple oscillations it is difficult to extract the cyclotron mass from the FFT spectra. Inspired by recent work\cite{veyrat_band_2015}, we can extract the cyclotron mass by studying the temperature dependence of the peak amplitudes in the oscillations via the Lifshitz-Kosevich formalism. The result is shown in Fig.~\ref{fig:AngularAndTemperature}d where we can study the evolution of the cyclotron mass upon varying the magnetic field where different oscillations contribute. Comparing this result with the FFT spectrum evolution in Fig.~\ref{fig:MRandOscillations}d in which we observe a single channel up to 0.07 T$^{-1}$, we can conclude that the channel corresponding to the $\alpha$ peak has a cyclotron mass $m_{\textup{c}}=(0.15\pm0.01)m_{\textup{e}}$, which is a typical value for the bulk conduction band\cite{kohler_conduction_1973}. Below 0.07 T$^{-1}$, we find a strong increase in the cyclotron mass up to $\raise.17ex\hbox{$\scriptstyle\sim$}0.28m_{\textup{e}}$ after which it lowers to $(0.20\pm0.01)m_{\textup{e}}$ and saturates. This higher value of $m_{\textup{c}}$ is probably due to the topological surface states\cite{wu_high-resolution_2015}, whereas a trivial 2DEG is supposed to have a similar mass as the bulk\cite{bianchi_coexistence_2010}. The interplay of the different oscillations could give rise to an increase in the cyclotron mass because channels with lower mobility ($\propto$ 1/$m_{\textup{c}}$) start contributing, provided that the scattering times in the different channels are the same\cite{veyrat_band_2015}. 

From the considerations above, we can match the charge-carrier densities extracted from the oscillations and from the magnetoresistance. From the FFT spectrum progression analysis, we can conclude that the $\alpha$ peak makes up the high mobility channel where the charge-carrier density $n_{\alpha}$ = (1.26$\pm$0.06)$\times10^{19}$cm$^{-3}$ when assuming bulk states ($n_{\textup{3D}} = k_{\textup{F,b}}^2k_{\textup{F,c}}/3\pi^2$) with an ellipsoid pocket with ellipticity $k_{\textup{F,c}}$/$k_{\textup{F,b}}$ = 1.8\cite{kulbachinskii_conduction-band_1999}. The value for $n_{\alpha}$ is in reasonable agreement with $n_{\textup{1}}/t$ found from the magnetoresistance analysis (Table.~\ref{tab:DensityMobility}). Furthermore, due to reduced scattering compared to that at any of the surfaces it is most likely that the high mobility channel corresponds to the unaffected bulk layer.

The state indicated by $\beta$ appears above 15 T and thus it is conceivable that this state is linked to the low mobility channel with $n_{\textup{2}}$. The origin of the observed 2D surface state, trivial or nontrivial, cannot be concluded from the determination of the Berry phase (see Supplemental Material below), but the extracted $m_{\textup{c}}$ at high fields hints at a topological surface state. Furthermore, it is not clear whether this state resides at the top or bottom surface because the characteristics of the electrostatics at both surfaces are unknown, which would affect the mobility. Assuming $n_{\textup{TSS}}$ = $k_{\textup{F}}^2/4\pi$ for a topological surface state and $n_{\textup{2DEG}}$ = $k_{\textup{F}}^2/2\pi$ for a possible 2DEG, the charge-carrier density related to $f_{\beta}$ varies between $n_{\textup{$\beta$,TSS}}=(6.7\pm0.3)\times10^{12}$cm$^{-2}$ and $n_{\textup{$\beta$,2DEG}}=(1.34\pm0.05)\times10^{13}$cm$^{-2}$, which makes up for 20 or 40 $\%$ of $n_{\textup{2}}$. We are careful to assume that this oscillation is linked to one surface state since it has been earlier reported that similar $n_{\beta}$ is present at the opposite surface\cite{brahlek_emergence_2014}, provided the mobilities at both surfaces are similar. Furthermore, as will be shown for $t$ = 20 QL, an additional peak between $\alpha$ and $2\alpha$ occurs which shows that additional states exist, which adds to the low mobility charge-carrier density $n_{\textup{2}}$.

\begin{figure}[t]
	\centering
		\includegraphics[width=\columnwidth]{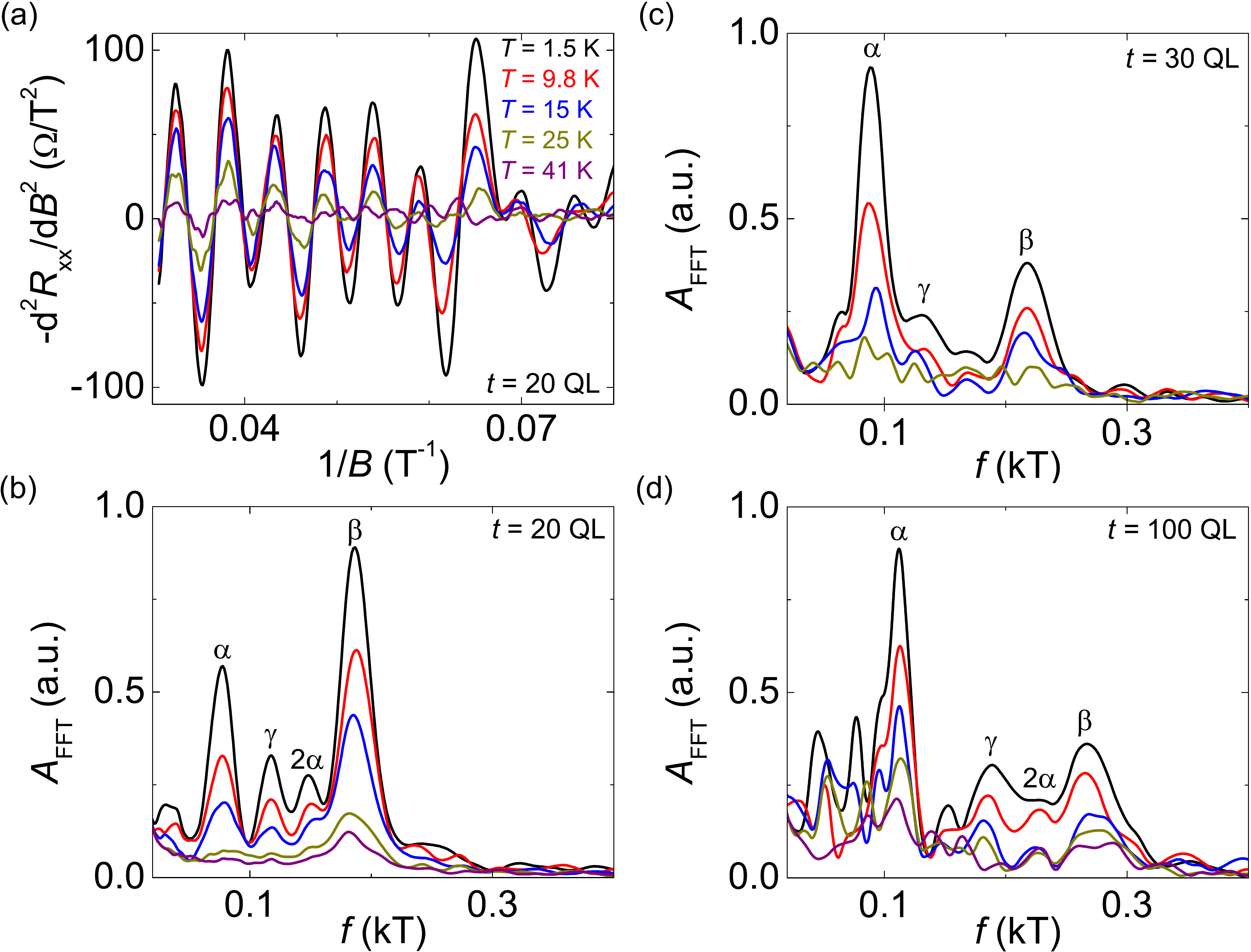}
		\caption{(a) The second derivative of the resistance with respect to the magnetic field --$d^2R_{\textup{xx}}$/$dB^2$ plotted vs 1/$B$ for $t$ = 20 QL. A clear oscillatory pattern is present with multiple oscillations. (b) Temperature dependence of the FFT spectrum (smoothed) based on the oscillations in (a). Colors correspond to the temperatures as depicted in (a). Temperature-dependent FFT spectra (smoothed) for (c) $t$ = 30 QL and (d) $t$ = 100 QL.}
	\label{fig:SpectraOtherThicknesses}
\end{figure}

The picture based on the charge-carrier densities for $t$ = 10 QL also applies for the samples with $t$ = 20 and 30 QL, but the correction for the ellipsoidal asymmetry is most probably smaller compared to the sample with $t$ = 10 QL which can be related to a lower charge-carrier density\cite{kulbachinskii_conduction-band_1999}. Comparing the two films with $t$ = 10 and 20 QL, we observe oscillations [Fig.~\ref{fig:SpectraOtherThicknesses}a] with a similar spectrum but with the presence of an additional $\gamma$ peak [Fig.~\ref{fig:SpectraOtherThicknesses}b], which could be a signature of a state at the surface opposite to where the channel linked to $\beta$ resides. Furthermore, the peak positions have changed, which is due to differences in charge-carrier density as also observed in the magnetoresistance measurements. From the similarities between these two samples, we can conclude that the thickness (i.e., bulk size) does not play a role but it is rather the relative mobilities and charge-carrier densities in these samples which are the decisive factors for the relative channel contributions. 

For the thicker samples, as shown in FIGs.~\ref{fig:SpectraOtherThicknesses}c and d, we observe a dominant peak $\alpha$ in the spectrum while the $2\alpha$, $\beta$, and $\gamma$ peaks are present but with a poor resolution. The reason for the decrease in amplitude is a lower signal-to-noise level of the measured voltage which generates a background and gives rise to a larger spectral width in the FFT spectrum. Furthermore, the oscillations show a beating pattern where oscillations of different frequencies partially cancel each other, yielding a loss of FFT amplitude. For the sample with $t$ = 100 QL, we find a poor agreement between the charge-carrier densities from the magnetoresistance and the SdH oscillations where the bulk state ($\alpha$ channel) can alone account for the total charge-carrier density $n_{\textup{1}}+n_{\textup{2}}$. The fitting of the magnetoresistance data shows that the mobilities and charge-carrier densities could be different from the extracted values as stronger oscillations were expected for the mobilities extracted. The difference between the fit and data could originate from additional channels with a more distinct mobility suggesting that the two-channel model is too limited to describe the data properly. Lastly, from atomic force microscopy (AFM) images (see Supplemental Material below) we observe height variations across the film surface, which might influence fitting parameters such as the effective thickness of the transport channel and can cause changes to the bulk density $n_{\textup{1}}/t$ on the order of $2\times10^{17}$cm$^{-3}$. This is in the same range of charge-carrier densities found for the 2D states. 

In conclusion, we find a good agreement between magnetoresistance data and the analysis of the SdH oscillations for Bi$_2$Se$_3$ thin films based on the extracted charge-carrier densities, where the channel contributions are quite unrelated to film thickness. We find that the bulk channel has a high mobility and is characterized by an ellipsoid Fermi pocket but a clear saturation in the angular dependence is absent. Due to the strong $g$ factor in these materials we observe a Zeeman splitting in our oscillations which has been observed before in optical measurements and investigations on thermoelectric effects under high magnetic field. Furthermore, we observe a pronounced 2D state, either topologically trivial or nontrivial, which partially accounts for the low mobility channel's charge-carrier density. Additional 2D states are observed but are often masked by the limited resolution of our analysis originating from the channel mobilities and charge-carrier densities. The limited resolution of the angular dependence and the difficulties to extract parameters like the Berry phase make it difficult to make a definitive statement on the origin of these states. 

The authors would like to thank L. Tang (Radboud University) for useful discussions on the processing of the data. Furthermore, the authors would like to thank J. G. Holstein, H. M. de Roosz, and T. J. Schouten (University of Groningen) for the technical support. We acknowledge the support of the HFML, member of the European Magnetic Field Laboratory. Work at the University of Groningen is supported by a Dieptestrategie grant from the Zernike Institute for Advanced Materials. Work at Rutgers University is supported by NSF (EFMA-1542798) and Gordon and Betty Moore Foundation's EPiQS Initiative (GBMF4418).


\clearpage

\onecolumngrid

\begin{center}
\textbf{Supplemental Material: Coexistence of bulk and surface states probed by Shubnikov--de Haas oscillations in Bi$_2$Se$_3$ with high charge-carrier density}
\end{center}

In this Supplemental Material additional data can be found for all the samples including the extracted charge-carrier densities from the SdH oscillations which can be compared with the values extracted from the Drude modelling. Furthermore, AFM images will be shown which give an idea about the growth quality and the error in the thickness leading to the error in $n_{\textup{1}}/t$. For $t$ = 10 QL, we also performed a Berry phase analysis. We will start out with details on the analysis performed in the main text.

\section*{S1. Details on analysis}

In this section we would like to clarify some of the analysis methods used in the main text. First of all, we would like to extend on the simultaneous fitting of the two-channel Drude model. From Equation (1) in the main text, we can find the expressions for the resistivity components by inverting the conductivity matrix:

\begin{subequations}
\label{eq:FinalRhoXXandXY}
\begin{gather}
R_{\textup{xx}}=\frac{\left(\frac{n_{\textup{1}}e\mu_{\textup{1}}}{1+\left(\mu_{\textup{1}} B\right)^{2}}+\frac{n_{\textup{2}}e\mu_{\textup{2}}}{1+\left(\mu_{\textup{2}} B\right)^{2}}\right)}{\left(\frac{n_{\textup{1}}e\mu_{\textup{1}}}{1+\left(\mu_{\textup{1}} B\right)^{2}}+\frac{n_{\textup{2}}e\mu_{\textup{2}}}{1+\left(\mu_{\textup{2}} B\right)^{2}}\right)^{2}+\left(\frac{n_{\textup{1}}e{\mu_{\textup{1}}}^{2}B}{1+\left(\mu_{\textup{1}} B\right)^{2}}+\frac{n_{\textup{2}}e{\mu_{\textup{2}}}^{2}B}{1+\left(\mu_{\textup{2}} B\right)^{2}}\right)^{2}}\\
R_{\textup{xy}}=\frac{\left(\frac{n_{\textup{1}}e{\mu_{\textup{1}}}^{2}B}{1+\left(\mu_{\textup{1}} B\right)^{2}}+\frac{n_{\textup{2}}e{\mu_{\textup{2}}}^{2}B}{1+\left(\mu_{\textup{2}} B\right)^{2}}\right)}{\left(\frac{n_{\textup{1}}e\mu_{\textup{1}}}{1+\left(\mu_{\textup{1}} B\right)^{2}}+\frac{n_{\textup{2}}e\mu_{\textup{2}}}{1+\left(\mu_{\textup{2}} B\right)^{2}}\right)^{2}+\left(\frac{n_{\textup{1}}e{\mu_{\textup{1}}}^{2}B}{1+\left(\mu_{\textup{1}} B\right)^{2}}+\frac{n_{\textup{2}}e{\mu_{\textup{2}}}^{2}B}{1+\left(\mu_{\textup{2}} B\right)^{2}}\right)^{2}}
\end{gather}
\end{subequations}

Here, $R_{\textup{xx}}$ is the (2D) sheet resistance and $R_{\textup{xy}}$ the Hall resistance. Simultaneous fitting is done via Matlab R2016a by minimalizing the sum of errors between data and fit of both $R_{\textup{xx}}$ and $R_{\textup{xy}}$ without any weighing. Here, the weak antilocalization (WAL) feature observed close to zero field does not affect the fitting procedure since the range over which this is observed is less than 1 $\%$ of the total field range. Furthermore, at elevated temperatures, where WAL is absent, the Drude model also shows a good agreement with the data and therefore we can rule out any fitting errors due to the presence of WAL. In this way, the fitting is most reliable due to inclusion of the full data set at once. 

Furthermore, we would like to elaborate on the analysis procedure for the Shubnikov--de Haas oscillations. To decouple the oscillations from the strong background we have taken the second derivative --$d^{2}R_{\textup{xx}}/$$dB^2$ since from Equation~\ref{eq:FinalRhoXXandXY} we have a $B^{2}$-dependence on the resistance in the limit of high field. By taking the second derivative, we are indeed successful to remove the background completely, whereas taking the first derivative gives a strong residual background where FFT analysis is difficult. 

Taking the second derivative requires a low noise level which we can decrease by adjacent averaging of the data. The averaging procedure is performed over 0.5 T intervals which are much shorter compared to the oscillation period such that this will not affect the FFT analysis; it only will slightly change the oscillation amplitude. Furthermore, the second derivative requires equidistant intervals such that interpolation is needed where the number of points is kept constant with respect to the original data. This interpolation is also used before the FFT is taken (in the 1/$B$ range) which does not yield any artifacts because of the large oscillation period in these measurements. 

\section*{S2. Additional data for sample with $t$ = 10 QL}

\begin{figure}
	\centering
		\includegraphics[width=\columnwidth]{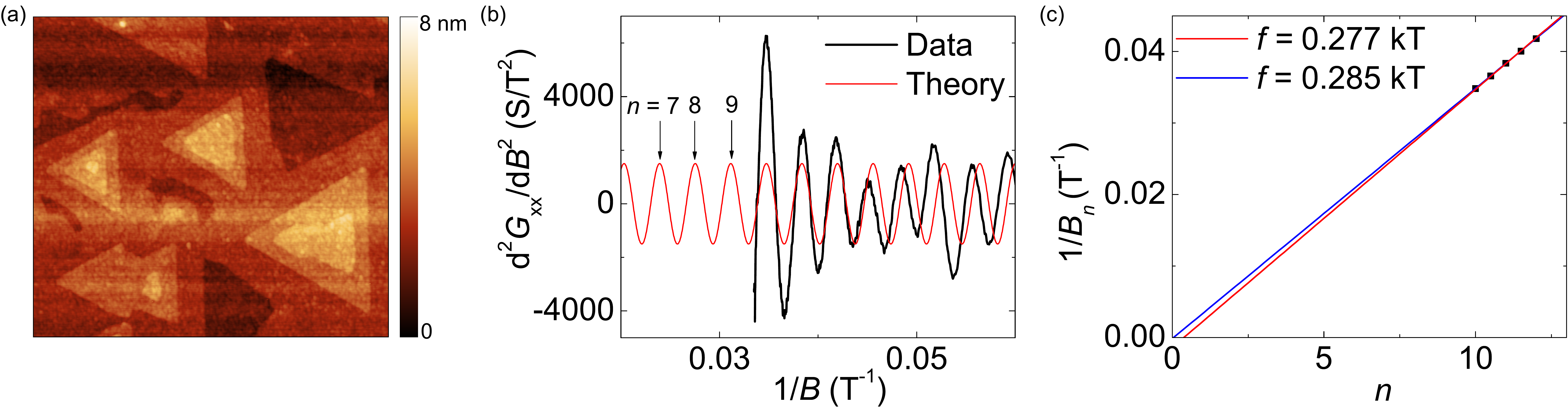}
		\caption{(a) AFM image (2$\times$2 $\micro$m$^2$) for sample with $t$ = 10 QL. (b) $d^{2}G_\textup{xx}$/$dB^{2}$ plotted vs 1/$B$ for Berry phase analysis (black). The red curve shows the theoretical positions of maxima and minima in $d^{2}G_\textup{xx}$/$dB^{2}$ with $f$ = 0.277 kT. (c) Landau level fan diagram extracted from the maxima and minima in the oscillatory pattern as shown in (b).}
	\label{fig:S1}
\end{figure}

A typical AFM image for this sample is shown in Fig.~\ref{fig:S1}a from which we assign a maximum uncertainty in the channel thickness $t$ of 1 QL as a conservative margin. 

To understand the origin of the 2D states, we could extract the Berry phase from the oscillations. A careful analysis has been described in \cite{ando_topological_2013} on how to index the maxima and minima in $dG_\textup{xx}$/$dB$. For our analysis, we looked at the maxima and minima in $d^{2}G_\textup{xx}$/$dB^{2}$ where the maxima in $d^{2}G_\textup{xx}$/$dB^{2}$ coincide with the minima in $G_\textup{xx}$ (labeled with integer $n$) and minima in $d^{2}G_\textup{xx}$/$dB^{2}$ correspond to maxima in $G_\textup{xx}$ (labeled $n + 1/2$ ($n$ = 1, 2, 3, ..)). From Fig.~\ref{fig:S1}b, it can be seen that the oscillation complies with the predicted behavior over a small range but due to the presence of multiple oscillations the data acquires a different phase compared to the theoretical curve. Therefore, the analysis can only be employed for the first three maxima. Plotting 1/$B_n$ versus $n$ yields a phase between 0.04 and 0.38 depending on the frequency taken; $f_{\beta}$ = 0.277 kT is the value found from the FFT analysis, $f_{\beta}$ = 0.285 kT corresponds to the best fit of the data and is within the error for $f_{\beta}$ as given in Table~\ref{tab:Density10QL}. In conclusion, due to the presence of multiple oscillations with different frequency, the limited number of oscillations and the error in $f_{\beta}$ a proper extraction of the Berry phase value cannot be made. 

At last, the values for $n_{\textup{TSS}}$ = $k_{\textup{F,b}}^2$/4$\pi$, $n_{\textup{2DEG}}$ = $k_{\textup{F,b}}^2$/2$\pi$, and $n_{\textup{3D}}$ = $k_{\textup{F,b}}^2k_{\textup{F,c}}$/3$\pi^2$ that we extract from the frequencies in the FFT spectra are displayed in Table~\ref{tab:Density10QL}. Here, we assume $k_{\textup{F,c}}/k_{\textup{F,b}}$=1.8 which has been reported for samples with similar charge-carrier density by Kulbachinskii et al.\cite{kulbachinskii_conduction-band_1999}. These values can be compared to $n_{\textup{1}}$ and $n_{\textup{2}}$ which are calculated for the 2D and 3D case and also included in Table~\ref{tab:Density10QL}. As also discussed in the main text, we observe a reasonable agreement between $n_{\alpha}$ (with $f$ = 0.115 kT) and $n_{\textup{1}}/t$. It is important to realize that when considering all channels to be 2D, the charge-carrier density from the oscillations clearly underestimates the values found from the two channel fit; the presence of a bulk channel therefore seems feasible and in agreement with the presence of a residual oscillation at $\theta$ = 90$^{\circ}$ and a Zeeman-split bulk state. The contribution of $n_{\beta}$ to $n_{\textup{2}}$ is 20 $\%$ when this channel is linked to a topological surface state. From band structure calculations\cite{zhang_topological_2009}, we find that the next bulk band is located around 1 eV higher than the first conduction band minimum which makes the origin of the state $\beta$ to be related to a second bulk band unlikely. In addition, the FFT spectra that we observe in comparison with the work by Kulbachinskii et al.\cite{kulbachinskii_conduction-band_1999} are different such that a second bulk state is more improbable. The presence of additional channels with an even lower mobility could account for the difference between the values extracted from the magnetoresistance and those from the SdH oscillations. 

\begin{table}
	\centering
	{
		\begin{tabular}{|M{1.2cm}|M{2cm}|M{3.5cm}|M{2.7cm}|M{2.5cm}|}
		\hline
		Label&$f$ (kT)&$n_{\textup{TSS}}$, $n_{\textup{2D}}$($\times$10$^{12}$cm$^{-2}$)\rule{0pt}{2.6ex}&$n_{\textup{2DEG}}$($\times$10$^{12}$cm$^{-2}$)&$n_{\textup{3D}}$($\times$10$^{19}$cm$^{-3}$)\\\hline
		$\alpha$&0.115$\pm$0.005&2.8$\pm$0.1&5.6$\pm$0.2&1.26$\pm$0.07\\\hline
		$\beta$&0.28$\pm$0.01&6.7$\pm$0.3&13.4$\pm$0.5&4.7$\pm$0.2\\\hline
		$n_{\textup{1}}$&&18$\pm$1&&1.8$\pm$0.1\\\hline
		$n_{\textup{2}}$&&34$\pm$2&&3.4$\pm$0.2\\\hline
		\end{tabular}}
	\caption{Extracted values for $n_{\textup{TSS}}$, $n_{\textup{2DEG}}$, and $n_{\textup{3D}}$ (with $k_{\textup{F,c}}/k_{\textup{F,b}}$ = 1.8) from the oscillations and $n_{\textup{1}}$ and $n_{\textup{2}}$ from the magnetoresistance (as can also be found in Table 1 of the main paper) for $t$ = 10 QL.}
	\label{tab:Density10QL}
\end{table}

\section*{S3. Data for sample with $t$ = 20 QL}

\begin{figure}
	\centering
		\includegraphics[width=\columnwidth]{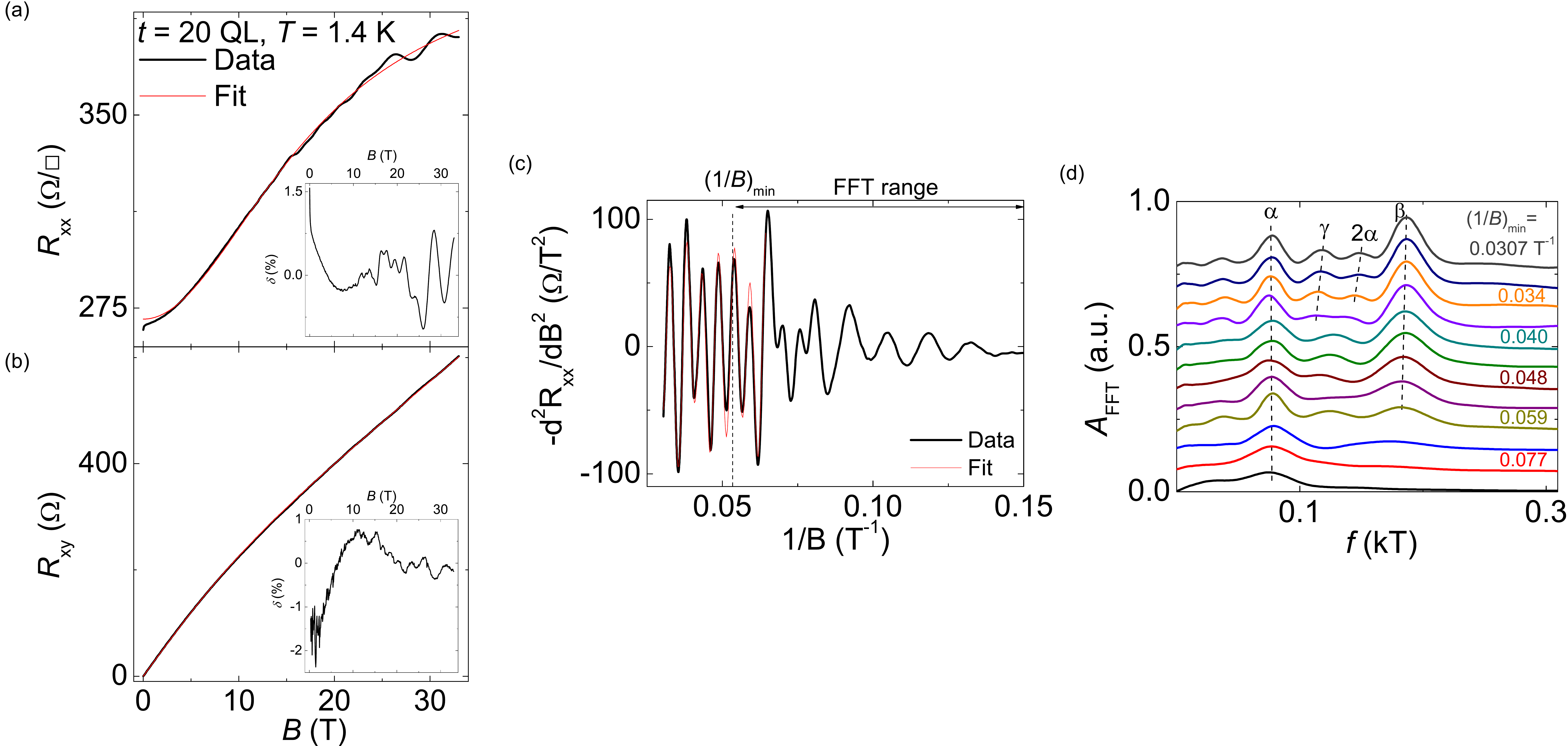}
		\caption{(a) Out-of-plane magnetic field dependence of (a) the longitudinal sheet resistance $R_{\textup{xx}}$ and (b) the Hall resistance $R_{\textup{xy}}$ for $t$ = 20 QL at $T$ = 1.4 K. The data (black) can be fitted with the two channel model (red) in good agreement as can be seen in the insets showing the residuals $\delta$ as defined in the main text. Oscillations in $R_{\textup{xx}}$ are again clearly visible beyond 15 T. (b) The second derivative of the resistance with respect to the magnetic field --$d^2R_{\textup{xx}}$/$dB^2$ plotted vs 1/$B$. Beyond 16 T (0.062 T$^{-1}$) the oscillatory pattern can be reconstructed from oscillations with $f_{\alpha}$ = 0.078$\pm$0.004 kT, $f_{\gamma}$ = 0.118$\pm$0.004 kT, and $f_{\beta} = $0.188$\pm$0.001 kT. (d) Magnetic field evolution of the (smoothed) FFT spectrum analyzed for different FFT ranges starting from 11 T towards higher fields with steps of 2 T, as depicted in (c). The FFT amplitude $A_{\textup{FFT}}$ is plotted vs frequency $f$(T) where the curves are offset by 0.07 for clarity.}
	\label{fig:S2}
\end{figure}

In Fig.~\ref{fig:S2}a and b the data and fits for the out-of-plane field dependence of the sheet resistance $R_{\textup{xx}}$ and the Hall resistance $R_{\textup{xy}}$, respectively, for sample with $t$ = 20 QL are shown where a good agreement between data and fit is observed. Importantly, for this sample the resistance has been measured for fields up to 33 T in order to include the last clear oscillation which enhances slightly the resolution in our FFT spectrum.   

When studying the FFT range progression analysis, as shown in Fig.~\ref{fig:S2}d, additional bands start to appear already below 15 T which is striking considering the low extracted mobility $\mu_{2}$ ($\textit{\micro}B\gg1$) and is inconsistent with the trend observed for the sample with $t$ = 10 QL. A direct assignment of the mobility to the appearance of the bands at certain fields is thus not straightforward. The larger amplitude of the oscillation contributes to a good resolution in FFT such that the spectral peaks already appear at lower fields. The order of appearance of the peaks as displayed in Fig.~\ref{fig:S2}d is similar to that described in the main text with the additional $\gamma$ peak starting to appear around similar fields as the harmonic peak. To reconstruct the oscillatory pattern, as shown in Fig.~\ref{fig:S2}c, $f_{\alpha}$, $f_{\beta}$, and $f_{\gamma}$ are found from the fit, whereas additional inclusion of $f_{\textup{2}\alpha}$ does not improve the fit much and only modifies the amplitude slightly.

\begin{figure}
	\centering
		\includegraphics[width=\textwidth]{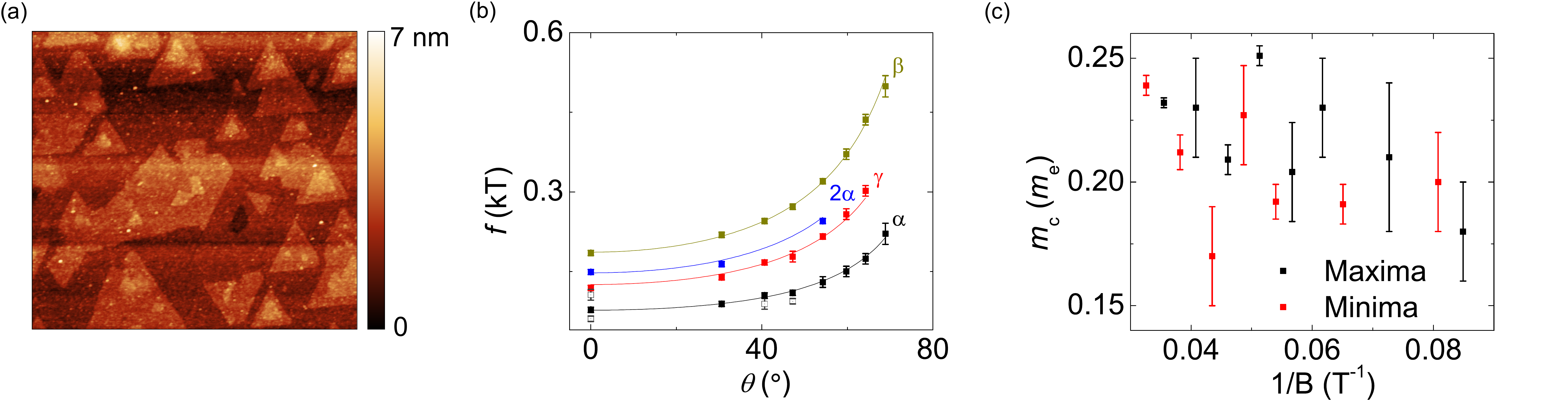}
		\caption{(a) AFM image (2$\times$2 $\micro$m$^2$) for sample with $t$ = 20 QL. (b) Angular dependence of the frequency extracted from the obtained FFT spectra. (c) Extracted cyclotron masses per peak position based on temperature dependence of the oscillations in main text [Fig. 3a].}
	\label{fig:S3}
\end{figure}

Table~\ref{tab:Density20QL} shows the extracted values for $n_{\textup{TSS}}$, $n_{\textup{2DEG}}$, and $n_{\textup{3D}}$ with $k_{\textup{F,c}}$$\approx$$k_{\textup{F,b}}$ as we have a lower charge-carrier density in this sample. Nevertheless, we can expect ellipticity as observed in the work by Kulbachinskii et al.\cite{kulbachinskii_conduction-band_1999}. These values can be compared to $n_{\textup{1}}$ and $n_{\textup{2}}$ as also displayed in Table~\ref{tab:Density20QL}. We find a good agreement between $n_{\alpha\textup{,3D}}$ and $n_{\textup{1}}/t$ giving good confidence on the origin of this state. As shown in Fig.~\ref{fig:S3}a, we observe hardly any thickness variation and therefore the error in $t$ is expected to be small. Furthermore, we find that the channel linked to $\beta$ makes up 20 $\%$ of $n_{\textup{2}}$, assuming it to be a topological surface state. The additional $\gamma$ state accounts for 13$\%$ of $n_{\textup{2}}$ which partially explains the difference in values found from the oscillations and the magnetoresistance. As shown in Fig.~\ref{fig:S3}, we observe hardly any thickness variation and therefore the error in $t$ is expected to be small. 

\begin{table}
	\centering
	{\begin{tabular}{|M{1.2cm}|M{2cm}|M{4.5cm}|M{2.7cm}|M{2.5cm}|}
		\hline
		Label&$f$ (kT)&$n_{\textup{TSS}}$, $n_{\textup{2D}}$($\times$10$^{12}$cm$^{-2}$)\rule{0pt}{2.6ex}&$n_{\textup{2DEG}}$($\times$10$^{12}$cm$^{-2}$)&$n_{\textup{3D}}$($\times$10$^{19}$cm$^{-3}$)\\\hline
		$\alpha$&0.077$\pm$0.002&1.86$\pm$0.05&3.7$\pm$0.1&0.38$\pm$0.02\\\hline
		$\gamma$&0.125$\pm$0.007&3.0$\pm$0.2&6.0$\pm$0.4&0.79$\pm$0.07\\\hline
		$\beta$&0.186$\pm$0.005&4.5$\pm$0.2&9.0$\pm$0.3&1.43$\pm$0.06\\\hline
		$n_{\textup{1}}$&&13$\pm$1&&0.67$\pm$0.03\\\hline
		$n_{\textup{2}}$&&24$\pm$2&&1.1$\pm$0.3\\\hline
		\end{tabular}}
	\caption{Extracted values for $n_{\textup{TSS}}$, $n_{\textup{2DEG}}$, and $n_{\textup{3D}}$ (with $k_{\textup{F,c}}$$\approx$$k_{\textup{F,b}}$) and $n_{\textup{1}}$ and $n_{\textup{2}}$ from the magnetoresistance (as can also be found in Table 1 of the main paper) for $t$ = 20 QL.}
	\label{tab:Density20QL}
\end{table}

Furthermore, we display the angular dependence for the sample with $t$ = 20 QL in where we see a similar behavior as for the $t$ = 10 QL sample, i.e. all the four peaks follow a 1/$\cos\theta$ dependence up to the angles where we are able to observe oscillations [Fig.~\ref{fig:S3}b]. For this sample, it is not possible to observe any remaining bulk oscillations at 90$^{\circ}$. 
At last, we show the cyclotron mass extracted per peak from the temperature dependence as also shown in Fig. 3a of the main paper [Fig.~\ref{fig:S3}c]. Also here we observe an increase in $m_{\textup{c}}$ towards higher magnetic field but not as structurally as for the data for $t$ = 10 QL. Furthermore, the cyclotron mass seems to be slightly higher compared to the sample with $t$ = 10 QL.

\section*{S4. Data for sample with $t$ = 30 QL}

\begin{figure}[!h]
	\centering
		\includegraphics[width=\textwidth]{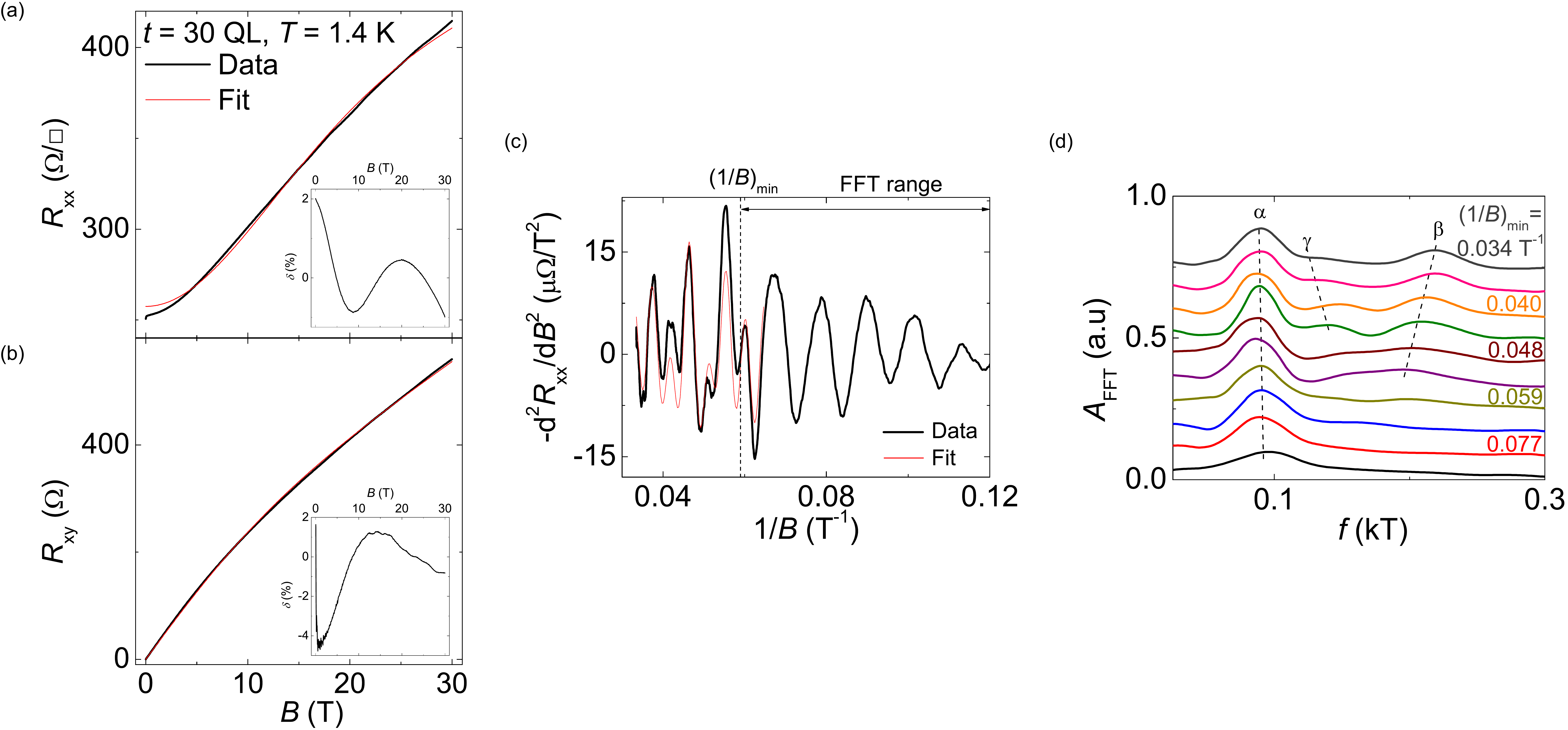}
		\caption{Out-of-plane magnetic field dependence of (a) the longitudinal sheet resistance $R_{\textup{xx}}$ and (b) the Hall resistance $R_{\textup{xy}}$ for $t$ = 30 QL at $T$ = 1.4 K. The data (black) can be fitted with the two channel model (red) in good agreement as can be seen in the insets showing the residuals $\delta$ as defined in the main text. (c) The second derivative of the resistance with respect to the magnetic field --$d^2R_{\textup{xx}}$/$dB^2$ plotted vs 1/$B$. Beyond 16 T (0.062 T$^{-1}$) the oscillatory pattern can be partially reconstructed from oscillations with $f_{\alpha}$ = 0.100$\pm$0.005 kT, $f_{\gamma}$ = 0.14$\pm$0.01kT, and $f_{\beta}$ = 0.219$\pm$0.005 kT, but a clear deviation from the data is present. (d) Magnetic field evolution of the (smoothed) FFT spectrum analyzed for different FFT ranges starting from 11 T towards higher fields with steps of 2 T, as depicted in (c). The FFT amplitude $A_{\textup{FFT}}$ is plotted vs frequency $f$(T) where the curves are offset by 0.08 for clarity.}
	\label{fig:S4}
\end{figure}

\begin{figure}[!h]
	\centering
		\includegraphics[width=\textwidth]{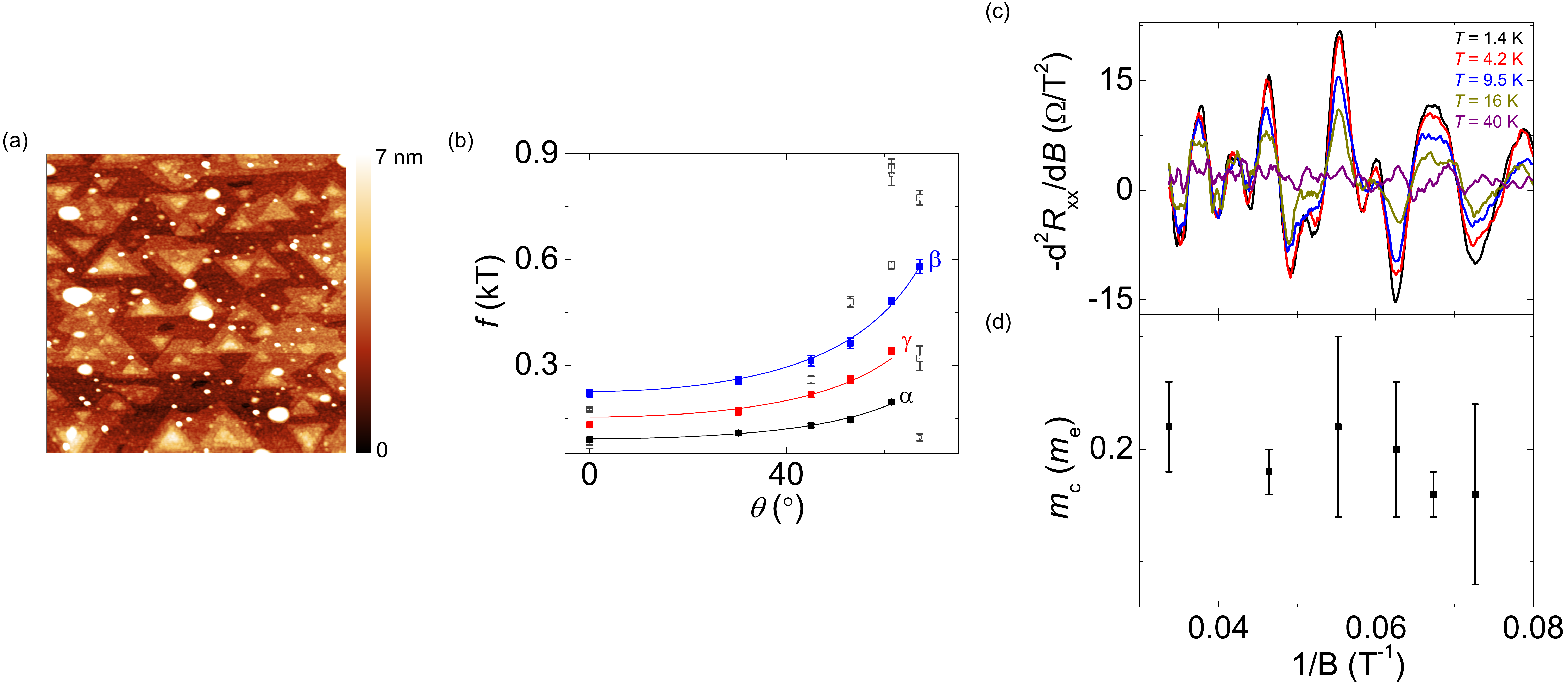}
		\caption{(a) AFM image (2$\times$2$\micro$m$^2$) of sample with $t$ = 30 QL from which we defined an error in the thickness of 1 QL. Clear contamination is visible. (b) Angular dependence of the frequency extracted from the obtained FFT spectra. (c) Temperature dependence of the oscillations. (d) Extracted cyclotron masses per peak position based on the oscillations in (c).}
	\label{fig:S5}
\end{figure}

In Fig.~\ref{fig:S4}a and b, the data and fits for the out-of-plane field dependence of the sheet resistance $R_{\textup{xx}}$ and the Hall resistance $R_{\textup{xy}}$, respectively, for the sample with $t$ = 30 QL are shown where a good agreement between data and fit is observed. In this case the SdH oscillations are not as clear as seen in the previous samples. Nevertheless, upon plotting --$d^2R_{\textup{xx}}$/$dB^2$ vs 1/$B$ we find a clear oscillatory pattern with an increased noise in the signals compared to the previously discussed samples [Fig.~\ref{fig:S4}c]. This increase in noise might be linked to the contamination that we find from the AFM image [Fig.~\ref{fig:S5}a]. We find that this oscillatory pattern shows three spectral peaks but not as clearly resolvable as previous samples where the 2$\alpha$ peak is missing [Fig.~\ref{fig:S4}d]. These three peaks can be used to fit the data, but a clear discrepancy is observed indicating that additional channels are present. Interestingly, we observe that the appearance of the peaks commences at rather low fields when considering the mobilities that we find from the magnetoresistivity fitting, similar to that observed for $t$ = 20 QL.

The presence of the peaks can also be checked by the angular dependence [Fig.~\ref{fig:S5}b]. Spectral peaks $\alpha$, $\beta$, and $\gamma$ show a clear angular dependence following a 1/cos$\theta$ behavior; again it is not possible to observe any remaining oscillation at $\theta$ = 90$^{\circ}$. Comparing the values extracted from the oscillations and the magnetoresistance [Table~\ref{tab:Density30QL}], we conclude that the $\alpha$ peak corresponds to the bulk channel with density $n_{\textup{1}}$. The other channels contribute to a total charge-carrier density of about (1.0$\pm$0.1)$\times10^{13}$cm$^{-2}$ which is on the same order of magnitude as $n_{\textup{2}}$, thereby assuming that these states are linked to topological surface states. From the cyclotron mass analysis [Fig.~\ref{fig:S5}d], we cannot see an evolution of $m_{\textup{c}}$ because of the large errors and the limited number of peaks that could be analyzed due to the presence of a beating pattern. 

\begin{table}[!htb]
	\centering
	{\begin{tabular}{|M{1.2cm}|M{2cm}|M{3.5cm}|M{2.7cm}|M{2.5cm}|}
		\hline
		Label&$f$ (kT)&$n_{\textup{TSS}}$, $n_{\textup{2D}}$($\times$10$^{12}$cm$^{-2}$)\rule{0pt}{2.6ex}&$n_{\textup{2DEG}}$($\times$10$^{12}$cm$^{-2}$)&$n_{\textup{3D}}$($\times$10$^{19}$cm$^{-3}$)\\\hline
		$\alpha$&0.092$\pm$0.006&2.2$\pm$0.2&4.5$\pm$0.3&0.50$\pm$0.05\\\hline
		$\gamma$&0.13$\pm$0.01&3.2$\pm$0.3&6.4$\pm0.5$&0.9$\pm$0.1\\\hline
		$\beta$&0.23$\pm$0.01&5.5$\pm$0.3&10.9$\pm$0.5&1.9$\pm$0.2\\\hline
		$n_{\textup{1}}$&&15$\pm$1&&0.49$\pm$0.02\\\hline
		$n_{\textup{2}}$&&27$\pm$1&&0.91$\pm$0.03\\\hline
		\end{tabular}}
	\caption{Extracted values for $n_{\textup{TSS}}$, $n_{\textup{2DEG}}$, and $n_{\textup{3D}}$ (with $k_{\textup{F,c}}$$\approx$$k_{\textup{F,b}}$) and $n_{\textup{1}}$ and $n_{\textup{2}}$ from the magnetoresistance (as can also be found in Table 1 of the main paper) for $t$ = 30 QL.}
	\label{tab:Density30QL}
\end{table}
\newpage

\section{S5. Data for sample with $t$ = 100 QL}

\begin{figure}[!h]
	\centering
		\includegraphics[width=\columnwidth]{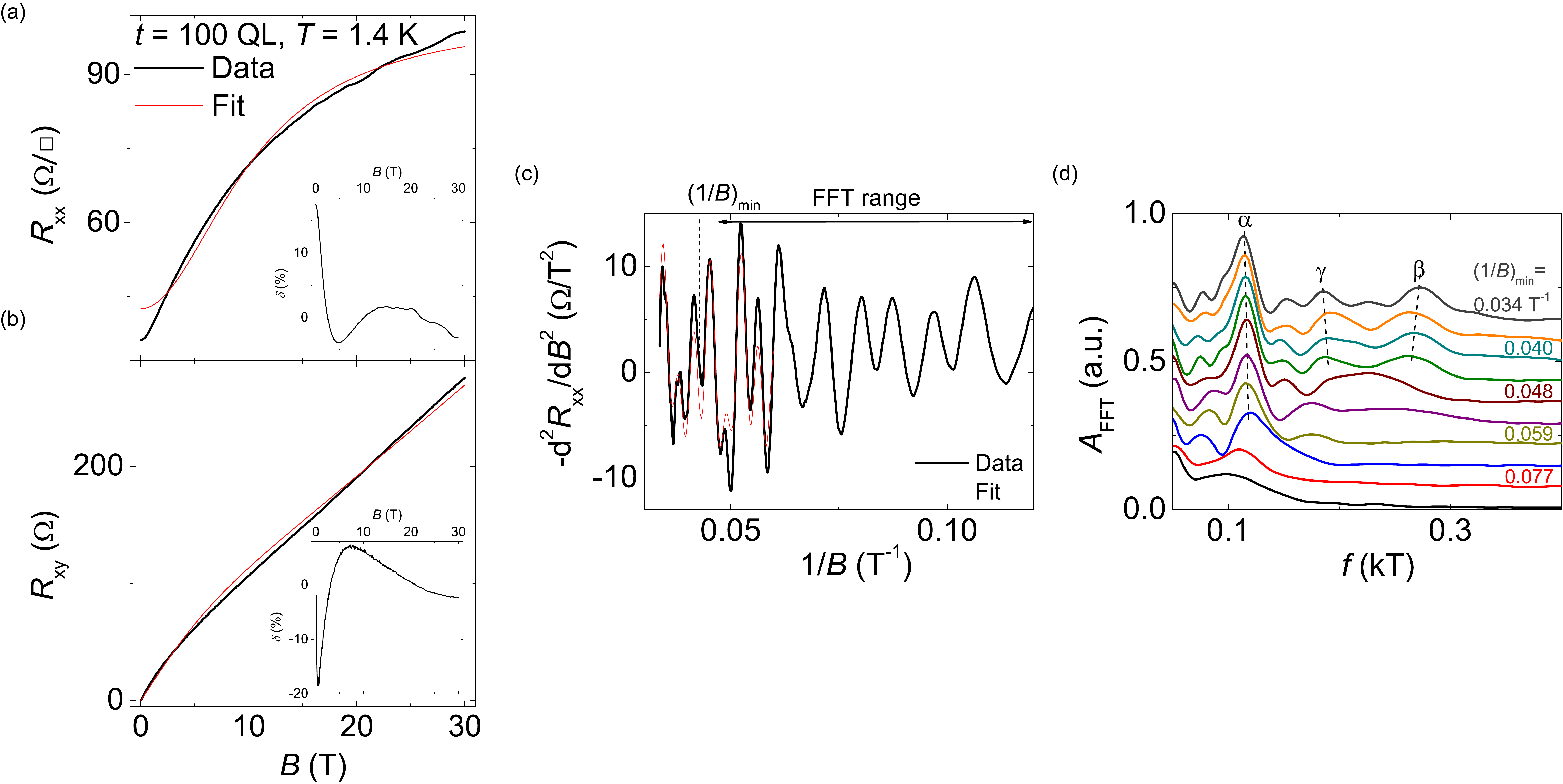}
		\caption{Out-of-plane magnetic field dependence of (a) the longitudinal sheet resistance $R_{\textup{xx}}$ and (b) the Hall resistance $R_{\textup{xy}}$ for $t$ = 100 QL at $T$ = 1.4 K. The data (black) and the two channel model fit (red) show disagreement, especially for $R_{\textup{xx}}$ as also can be seen in the insets showing the residuals $\delta$ as defined in the main text. (c) The second derivative of the resistance with respect to the magnetic field --$d^2R_{\textup{xx}}$/$dB^2$ plotted vs 1/$B$. Oscillations are clearly visible. Beyond 17 T (0.06 T$^{-1}$) the oscillatory pattern can partially be reconstructed from oscillations with $f$ = 0.110$\pm$0.005, 0.15$\pm$0.01, 0.175$\pm$0.005, and 0.272$\pm$0.005 kT. (d) Evolution of the (smoothed) FFT spectrum analyzed for different regions (i)-(iv) as depicted in (c) where the FFT amplitude $A_{\textup{FFT}}$ is plotted vs frequency $f$(T).}
	\label{fig:S6}
\end{figure}

\begin{figure}
	\centering
		\includegraphics[width=\textwidth]{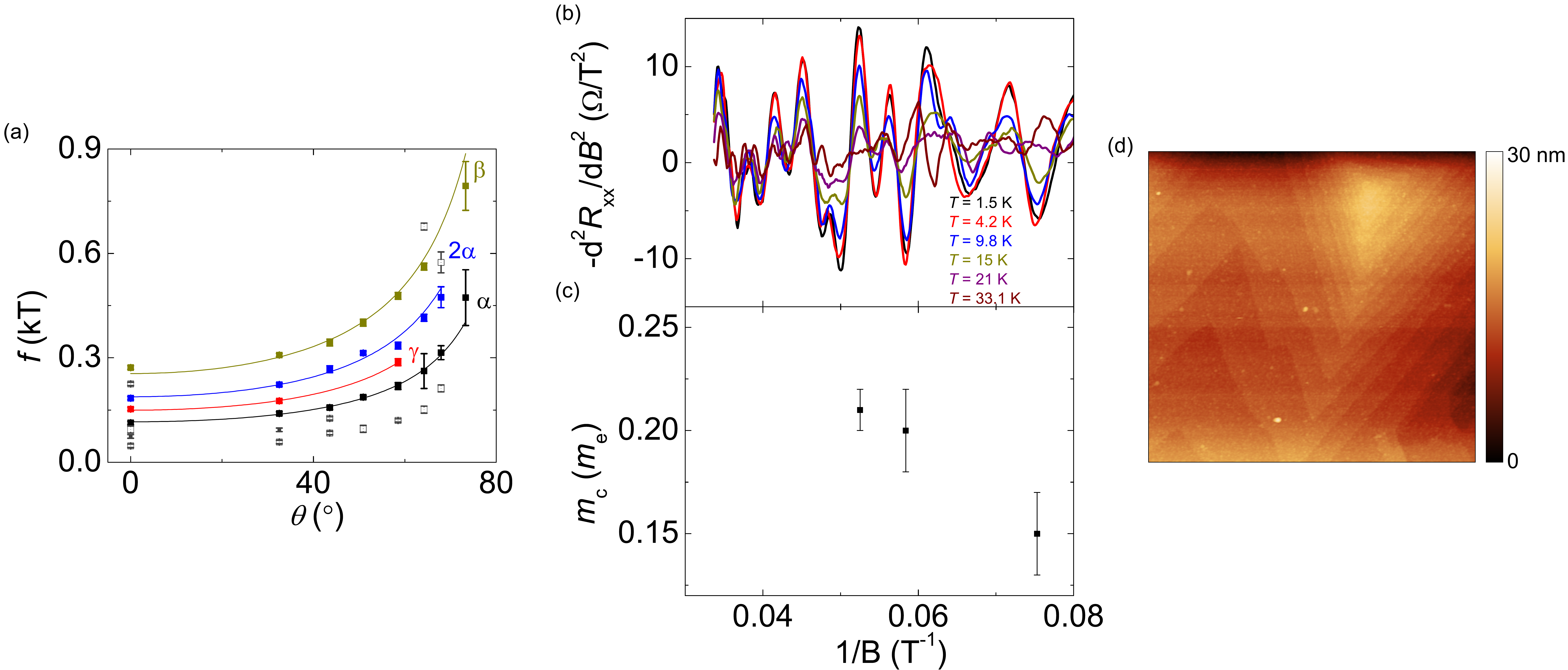}
		\caption{(a) Angular dependence of the frequency extracted from the obtained FFT spectra for $t$ = 100 QL. (b) Temperature dependence of the oscillations. (c) Extracted cyclotron masses per peak position based on oscillations (b). (d) AFM image (2$\times$2$\micro$m$^2$) of sample from where we defined an error in the thickness of 5 QL.}
	\label{fig:S7}
\end{figure}

In Fig.~\ref{fig:S6}a, the data and fits for the out-of-plane field dependence of the sheet resistance $R_{\textup{xx}}$ and the Hall resistance $R_{\textup{xy}}$, respectively, for the sample with $t$ = 100 QL are shown where there is a disagreement between data and fit, which means that for this thickness the two channel model is too limited to describe the magnetoresistance. Upon plotting --$d^2R_{\textup{xx}}$/$dB^2$ vs 1/$B$ we find a clear oscillatory pattern with an increased noise in the signals compared to the previously discussed samples [Fig.~\ref{fig:S6}c]. In FIG~\ref{fig:S6}d, we find that this oscillatory pattern contains three main spectral peaks and a weak harmonic 2$\alpha$ peak. To reconstruct the oscillatory pattern the four given frequencies are not sufficient enough which indicates more channels might be present but are beyond our resolution. A hint for an additional peak is given by the presence of an additional trace in the angular dependence as shown in Fig.~\ref{fig:S7}a at a frequency $f$ = 0.046 kT but was considered an artifact in the FFT analysis because of the unclear temperature dependence in Fig. 3d of the main paper. 

Due to the beating features in the oscillation pattern [Fig.~\ref{fig:S7}b], it is difficult to extract the cyclotron mass for this sample as shown in Fig.~\ref{fig:S7}c. Nevertheless, we seem to find a cyclotron mass $m_{\textup{c}}$ = 0.15$m_{\textup{e}}$ which increases to 0.20$m_{\textup{e}}$ after 15 T which is in agreement with the trend as observed for $t$ = 10 QL. At last, we show in Table~\ref{tab:Density100QL} the extracted charge-carrier densities where we observe that any extracted bulk value from the oscillations is larger than $n_{\textup{1}}/t$ and $n_{\textup{2}}/t$. As mentioned in the main text, the charge-carrier density from $f_{\alpha}$ = 0.12 kT could account for $n_{\textup{1}}$+$n_{\textup{2}}$. One uncertainty is the effective thickness $t$ of the sample as shown in the AFM image [Fig.~\ref{fig:S7}d], where there are large triangular undulations present at the surface. Nevertheless, the exact reason for the disagreement between magnetoresistance measurements and the SdH oscillations analysis is unclear.

\begin{table}
	\centering
	{\begin{tabular}{|M{1.2cm}|M{2cm}|M{3.5cm}|M{2.7cm}|M{2.5cm}|}
		\hline
		Label&$f$ (kT)&$n_{\textup{TSS}}$, $n_{\textup{2D}}$($\times$10$^{12}$cm$^{-2}$)\rule{0pt}{2.6ex}&$n_{\textup{2DEG}}$($\times$10$^{12}$cm$^{-2}$)&$n_{\textup{3D}}$($\times$10$^{19}$cm$^{-3}$)\\\hline
		$\alpha$&0.12$\pm$0.01&2.8$\pm$0.3&5.6$\pm$0.5&0.7$\pm$0.1\\\hline
		$\gamma$&0.188$\pm$0.003&4.55$\pm$0.08&9.1$\pm$0.2&1.46$\pm$0.04\\\hline
		$\beta$&0.25$\pm$0.02&6.1$\pm$0.5&12$\pm$1&2.3$\pm$0.3\\\hline
		$n_{\textup{1}}$&&27$\pm$2&&0.33$\pm$0.02\\\hline
		$n_{\textup{2}}$&&42$\pm$2&&0.42$\pm$0.02\\\hline
		\end{tabular}}
	\caption{Extracted values for $n_{\textup{TSS}}$, $n_{\textup{2DEG}}$, and $n_{\textup{3D}}$ (with $k_{\textup{F,c}}$$\approx$$k_{\textup{F,b}}$) and $n_{\textup{1}}$ and $n_{\textup{2}}$ from the magnetoresistance (as can also be found in Table 1 of the main paper) for $t$ = 100 QL.}
	\label{tab:Density100QL}
\end{table}
\end{document}